\documentclass{article}[10pt]
\usepackage[titletoc]{appendix}
\usepackage{latexsym}
\usepackage{amssymb}
\usepackage{amsmath}
\usepackage{color}
\usepackage{amsfonts,amssymb}
\usepackage{mathrsfs}
\usepackage{dsfont}
\usepackage{graphicx}
\usepackage{subfigure}
\usepackage{cite}
\usepackage[center]{titlesec}
\renewcommand{\theequation}
{\arabic{section}.\arabic{equation}}

\newtheorem{prop}{Proposition}
\newtheorem{algorithm}{Algorithm}
\newtheorem{thm}{Theorem}[section]
\newtheorem{rem}{Remark}
\begin{document}

\noindent{\bf Integrable Discretization of Soliton Equations via Bilinear Method and B\"acklund Transformation}

\noindent{Yingnan Zhang$^{1,2}$, Xiangke Chang$^{1,2}$, Juan Hu$^{3}$, \\
Xingbiao Hu$^{1}$ and Hon-Wah Tam$^{4}$}

\noindent\footnotesize{
\begin{tabular}{ l  p{12.2cm}}
$1$ &  LSEC, Institute of Computational Mathematics and Scientific Engineering Computing, Academy of Mathematics and System Sciences, Chinese Academy of Sciences, Beijing 100190,China\\
$2$ &  University of Chinese Academy of Sciences,Beijing, China\\
$3$ &  Department of Applied Mathematics, Zhejiang University of Technology, Hangzhou 310023, PR China\\
$4$ &  Department of Computer Science, Hong Kong Baptist University, Kowloon Tong, Hong Kong, China
\end{tabular}}\\[6mm]

\footnotetext{\hspace*{5.8mm}Emails: zhangyingnan@lsec.cc.ac.cn, changxk@lsec.cc.ac.cn, hujuan@lsec.cc.ac.cn, hxb@lsec.cc.ac.cn, tam@comp.hkbu.edu.hk}

\normalsize\noindent{ \bf{Abstract}}\quad
In this paper, we present a systematic procedure to derive discrete
analogues of integrable PDEs via Hirota's bilinear method.  This
approach is mainly based on the compatibility between an integrable
system and its B\"acklund transformation. We apply this procedure to
several equations, including the extended Korteweg-de-Vries (KdV) equation,
the extended Kadomtsev-Petviashvili (KP) equation, the extended Boussinesq equation,
the extended Sawada-Kotera (SK) equation and the extended Ito equation, and obtain
their associated semi-discrete analogues. In the continuum limit,
these differential-difference systems converge to their
corresponding smooth equations. For these new  integrable systems,
their B\"acklund transformations and Lax pairs are derived.\vspace{0.3cm}


\section{Introduction}

The integrable discretization of soliton equations or integrable
systems has attracted a lot of interest \cite{i1,i2} and has
been studied for many years from various viewpoints.
Although there
is no commonly accepted unified mathematical definition of integrable
systems, the most widely accepted features of these kinds of
equations include the zero curvature representation or Lax pair,
B\"acklund transformation and multi-soliton solutions.

The problem
of integrable discretization is to discretize an integrable
differential equation, meanwhile preserving its integrability.  Ablowitz and Ladik \cite{AL}, Hirota
\cite{hr1,hr2,hr3}, Nijhoff \cite{ncwq,qncl} and others did pioneer work on the
integrability of difference and differential-difference equations
over 30 years ago. As stated by Kakei Saburo in \cite{kakei},
besides the bilinear methods
\cite{hr1,hr2,hr3,djm1,djm2,djm3,djm4,djm5}, there are several other famous
procedures for integrable discretization, including\\
\begin{tabular}{ l  p{11cm}}
$\bullet$ &    discretization of Lax pairs \cite{AL}\\
$\bullet$ &    discretization based on the dressing method \cite{levi}\\
$\bullet$ &    the direct linearization method \cite{ncwq,qncl}\nonumber\\
$\bullet$ &    discretization based on the construction of Poisson structure \cite{FV,sur}.\\
\end{tabular}\\
Here we have only given a few representative references. In
addition, there is some recent progress in \cite{J1} where a method
based on loop group is introduced. Other recent works can be found in \cite{feng1, feng2, hy, feng3,suris1,suris2,
veni,yu1,yu2,zhang1,zhang2} and references there.

\setlength{\unitlength}{1mm}
\centerline{
\begin{picture}(100,125)
\put(5,110){
\newsavebox{\processa}
\savebox{\processa}(30,15)
{\put(0,15){\line(1,0){30}}
\put(0,0){\line(0,1){15}}
\put(0,0){\line(1,0){30}}
\put(30,0){\line(0,1){15}}
\put(8,10){integrable}
\put(8,6){differential}
\put(8,2){equation}}
\usebox{\processa}}
\put(7,105){variable}
\put(1,102){transformation}
\put(25,109){\vector(0,-1){8}}
\put(5,85){
\newsavebox{\processb}
\savebox{\processb}(30,15)
{\put(0,15){\line(1,0){30}}
\put(0,0){\line(0,1){15}}
\put(0,0){\line(1,0){30}}
\put(30,0){\line(0,1){15}}
\put(8,10){bilinear}
\put(8,6){differential}
\put(8,2){equation}}
\usebox{\processb}}
\put(4,80){discretization}
\put(25,84){\vector(0,-1){8}}
\put(5,60){
\newsavebox{\processc}
\savebox{\processc}(30,15)
{\put(0,15){\line(1,0){30}}
\put(0,0){\line(0,1){15}}
\put(0,0){\line(1,0){30}}
\put(30,0){\line(0,1){15}}
\put(5,10){bilinear}
\put(5,6){difference}
\put(5,2){equation$(eq1)$}}
\usebox{\processc}}
\put(5,55){integrability}
\put(25,59){\vector(0,-1){8}}
\put(5,35){
\newsavebox{\processd}
\savebox{\processd}(30,15)
{\put(0,15){\line(1,0){30}}
\put(0,0){\line(0,1){15}}
\put(0,0){\line(1,0){30}}
\put(30,0){\line(0,1){15}}
\put(5,10){bilinear}
\put(5,6){difference}
\put(5,2){equation$(eq2)$}}
\usebox{\processd}}
\put(7,30){variable}
\put(1,27){transformation}
\put(25,34){\vector(0,-1){8}}
\put(5,10){
\newsavebox{\processe}
\savebox{\processe}(30,15)
{\put(0,15){\line(1,0){30}}
\put(0,0){\line(0,1){15}}
\put(0,0){\line(1,0){30}}
\put(30,0){\line(0,1){15}}
\put(8,10){integrable}
\put(8,6){difference}
\put(8,2){equation}}
\usebox{\processe}}
\put(65,110){
\newsavebox{\processaa}
\savebox{\processaa}(30,15)
{\put(0,15){\line(1,0){30}}
\put(0,0){\line(0,1){15}}
\put(0,0){\line(1,0){30}}
\put(30,0){\line(0,1){15}}
\put(8,10){integrable}
\put(8,6){differential}
\put(8,2){equation}}
\usebox{\processaa}}
\put(67,105){variable}
\put(61,102){transformation}
\put(85,109){\vector(0,-1){8}}
\put(65,85){
\newsavebox{\processbb}
\savebox{\processbb}(30,15)
{\put(0,15){\line(1,0){30}}
\put(0,0){\line(0,1){15}}
\put(0,0){\line(1,0){30}}
\put(30,0){\line(0,1){15}}
\put(8,10){bilinear}
\put(8,6){differential}
\put(8,2){equation}}
\usebox{\processbb}}
\put(64,80){compatibility}
\put(68,76){of BT}
\put(85,84){\vector(0,-1){8}}
\put(65,60){
\newsavebox{\processcc}
\savebox{\processcc}(30,15)
{\put(0,15){\line(1,0){30}}
\put(0,0){\line(0,1){15}}
\put(0,0){\line(1,0){30}}
\put(30,0){\line(0,1){15}}
\put(5,10){expanded}
\put(5,6){bilinear}
\put(5,2){system$(eq3)$}}
\usebox{\processcc}}
\put(52,55){\textcircled{1}introduction of}
\put(54,52){discrete variable}
\put(87,55){\textcircled{2}convergence}
\put(89.5,52){condition}
\put(85,59){\vector(0,-1){8}}
\put(65,35){
\newsavebox{\processdd}
\savebox{\processdd}(30,15)
{\put(0,15){\line(1,0){30}}
\put(0,0){\line(0,1){15}}
\put(0,0){\line(1,0){30}}
\put(30,0){\line(0,1){15}}
\put(5,10){bilinear}
\put(5,6){difference}
\put(5,2){equation$(eq4)$}}
\usebox{\processdd}}
\put(67,30){variable}
\put(61,27){transformation}
\put(85,34){\vector(0,-1){8}}
\put(65,10){
\newsavebox{\processee}
\savebox{\processee}(30,15)
{\put(0,15){\line(1,0){30}}
\put(0,0){\line(0,1){15}}
\put(0,0){\line(1,0){30}}
\put(30,0){\line(0,1){15}}
\put(8,10){integrable}
\put(8,6){difference}
\put(8,2){equation}}
\usebox{\processee}}
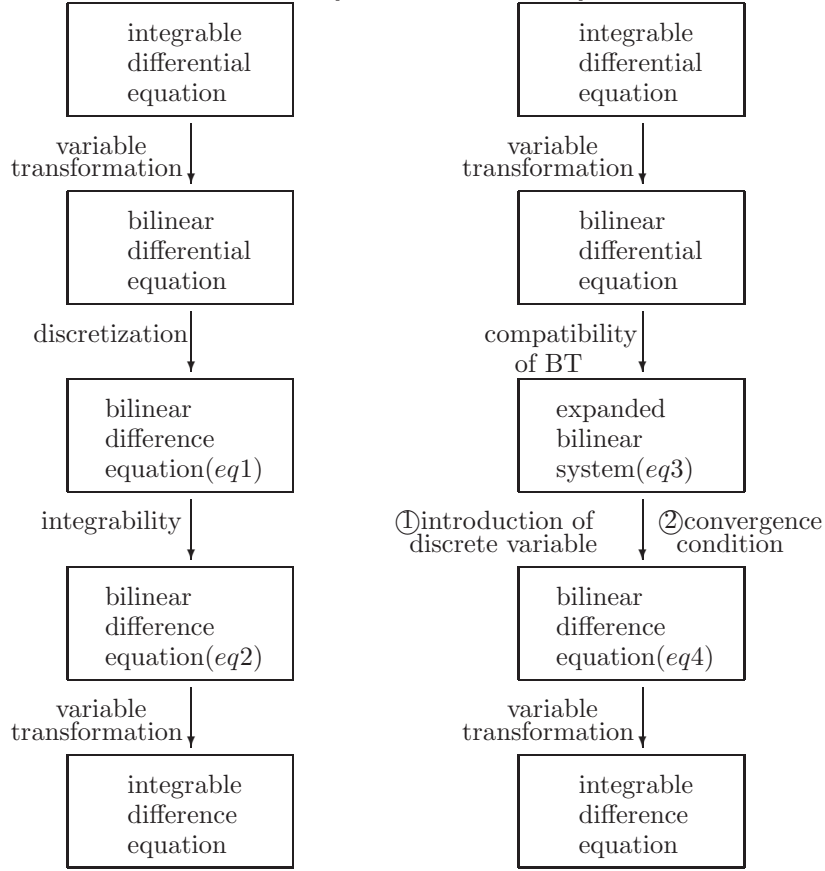
\put(1,5){Fig. 1: Flow charts of bilinear method of
integrable discretization --} \put(1,1){Left chart: traditional
approach; Right chart: new approach}
\end{picture}}

There is a well-known idea that integrable discretizations are provided
by a suitable interpretation of B\"acklund transformations.
Generating integrable differential-difference equations from B\"acklund transformations, as we know, was first studied by Chiu and Ladik in \cite{ladik} and Levi and Benguria in \cite{levi1,levi2}. Here in this paper, we showed that, by introducing a convergence condition, the integrable discretization of a soliton equation can be derived from the compatibility between the integrable differential equation and its B\"acklund transformation.

The procedure presented  here is from a different viewpoint of the bilinear method and is much more direct.
The traditional bilinear method \cite{hr1,hr4,h1,h2} as shown in the left chart of Fig. 1, is to
discretize the bilinear form of the soliton equation first and then
confirm the integrability. Due to non-uniqueness, there is some
freedom in ($eq1$) when we discretize the
smooth bilinear differential equation. Subsequently ($eq2$) is obtained from ($eq1$) by
considering its integrabilities (soliton solutions or B\"acklund transformation).
Different from the traditional approach, the new procedure shown on
the right of Fig. 1 preserves the integrability first and then
discretize the equation. Given an integrable bilinear equation and
its B\"acklund transformation, an expanded system ($eq3$) of
bilinear equations compatible with the original bilinear equation
can be obtained. In ($eq3$), there are some free parameters inherited
from the associated B\"acklund transformation. Using properties of
the $\tau$-function or the ideas in \cite{levi1}, it is natural to
introduce some discrete variables in the new integrable system. By
considering continuum limit, we impose a convergence condition on
the new integrable system. Then the parameters can be determined and
an integrable discretization $eq4$ of the original bilinear system
can be derived.
We will illustrate this  approach with the extended
Korteweg-de-Vries (KdV) system.

The extended KdV system is
given by
\begin{eqnarray}
&&v_t=\frac{1}{4}q+\frac{3}{2}u^2\label{kdv11},\\
&&u_t=\frac{1}{4}r+3up\label{kdv12},\\
&&p=u_x,q=p_x,r=q_x.\label{kdv13}
\end{eqnarray}
The dependent variable
transformation $v=(ln f)_{x}$, $u=(ln f)_{xx}$ produce the bilinear form
\begin{equation}
 D_x(D_t-\frac{1}{4}D_x^3)f \cdot f=0.\label{kdv2}
\end{equation}
Here the $D$-operator is defined by
\begin{eqnarray}
&& D_t^mD_x^n a(t,x)\cdot b(t,x)=\frac{\partial^m}{\partial s^m}\frac{\partial^n}{\partial y^n}a(t+s,x+y)b(t-s,x-y)|_{s=0,y=0},\nonumber\\
&& \qquad \qquad \qquad \qquad \qquad \qquad \qquad \qquad \qquad
m,n=0,1,2,\cdots,
\end{eqnarray}
or by the exponential identity
\begin{eqnarray}
&& exp( \delta D_z)a(z)\cdot b(z)=exp( \delta \partial_y)(a(z+y)b(z-y))|_{y=0},\nonumber\\
&& \qquad \qquad \qquad \qquad =a(z+\delta)b(z-\delta).
\end{eqnarray}
In the following, we will call the extended KdV system \eqref{kdv11}-\eqref{kdv13} and its bilinear form \eqref{kdv2} KdV equation without diffusion.
The following bilinear B\"acklund transformation of \eqref{kdv2} has
been given in \cite{d1}:
\begin{eqnarray}
&& (D_x^2-\lambda D_x)f \cdot g=0,\label{kdv-b1}\\
&& (D_t-\frac{1}{4}D_x^3+\mu)f \cdot g=0,\label{kdv-b2}
\end{eqnarray}
where $\lambda, \mu$ are arbitrary constants. For details of the
$D$-operator and the bilinear B\"acklund transformation,
see\cite{d1}. Here we consider \eqref{kdv2} and \eqref{kdv-b1}
together as a new system
\begin{eqnarray}
 && D_x(D_t-\frac{1}{4}D_x^3)f \cdot f=0,\\
 && (D_x^2-\lambda D_x)f \cdot g=0.
\end{eqnarray}
Taking  $f\rightarrow f_{n}$ and $g\rightarrow f_{n-h}$, where $n$ is a discrete variable and $h$ is the step size, we get a differential-difference system
\begin{eqnarray}
 && D_x(D_t-\frac{1}{4}D_x^3)f_n \cdot f_n=0,\label{kdv-c1}\\
 && (D_x^2-\lambda D_x)f_n \cdot f_{n-h}=0.\label{kdv-c2}
\end{eqnarray}
From properties of the B\"acklund transformation, the above system
is also integrable.

\begin{rem}
It will be better to write $f_{n+h}$ as $f_{n+1}$. For the sake of introducing a convergence condition $D_x=D_n+O(h)$, we keep using the form $f_{n+h}$.
\end{rem}

 To obtain the integrable discretization of \eqref{kdv2}, the convergence condition
$D_x=D_n+O(h^k)$, $k\geq1$ should be upheld. One can rewrite
\eqref{kdv-c2} as
\begin{equation}
[D_x^2\cosh(\frac{h}{2}D_n)-\lambda D_x\sinh(\frac{h}{2}D_n)]f_n\cdot f_n=0.
\end{equation}
Expanding this equation in powers of $h$, we get
\begin{equation}
[D_x^2-\frac{h}{2}\lambda D_xD_n+O(h^2)]f_n\cdot f_n=0.
\end{equation}
By imposing the convergence condition $D_x=D_n+O(h^k)$, $k\geq1$, we
have $\lambda=\frac{2}{h}$.
\begin{rem}
With the convergence condition $D_x=D_n+O(h^k)$, $k\geq1$, the
discrete variable $n$ can be viewed as an approximation to $x$ and
$h$ is just the step size in the $x$-direction. We do not discretize
$x$ directly but take it as an auxiliary variable. In the bilinear
equations, we still write it as $x$ without confusion.
\end{rem}

We next show that
\begin{eqnarray}
 && D_x(D_t-\frac{1}{4}D_x^3)f_n \cdot f_n=0,\label{kdv-d1}\\
 && (D_x^2-\frac{2}{h} D_x)f_n \cdot f_{n-h}=0,\label{kdv-d2}
\end{eqnarray}
is an integrable discretization of the KdV equation. Applying the dependent variable transformation
$v_n=(ln f_n)_{x}$, $ u_n=v_{n,x}$, $p_n=u_{n,x}$, $ q_n=p_{n,x}$, $r_n=q_{n,x}$ to \eqref{kdv-d1}-\eqref{kdv-d2}, we get
\begin{eqnarray}
&& u_{n,t}=\frac{1}{4}r_n+3u_np_n,\label{kdv-d3}\\
&& v_{n,t}=\frac{1}{4}q_n+\frac{3}{2}u_n^2,\label{kdv-d4}\\
&& (p_{n+\frac{h}{2}}+ p_{n-\frac{h}{2}})
=\frac{2}{h}(u_{n+\frac{h}{2}}-u_{n-\frac{h}{2}})-2(u_{n+\frac{h}{2}}-u_{n-\frac{h}{2}})(v_{n+\frac{h}{2}}-v_{n-\frac{h}{2}}),\label{kdv-d5}\\
&& (q_{n+\frac{h}{2}}+q_{n-\frac{h}{2}})
=\frac{2}{h}(p_{n+\frac{h}{2}}-p_{n-\frac{h}{2}})-2(p_{n+\frac{h}{2}}-p_{n-\frac{h}{2}})(v_{n+\frac{h}{2}}-v_{n-\frac{h}{2}})\nonumber\\
&&\qquad\qquad\qquad\qquad-2(u_{n+\frac{h}{2}}-u_{n-\frac{h}{2}})^2,\label{kdv-d6}\\
&& (r_{n+\frac{h}{2}}+r_{n-\frac{h}{2}})
=\frac{2}{h}(q_{n+\frac{h}{2}}-q_{n-\frac{h}{2}})-6(p_{n+\frac{h}{2}}-p_{n-\frac{h}{2}})(u_{n+\frac{h}{2}}-u_{n-\frac{h}{2}})\nonumber\\
&&\qquad\qquad\qquad\qquad-2(q_{n+\frac{h}{2}}-q_{n-\frac{h}{2}})(v_{n+\frac{h}{2}}-v_{n-\frac{h}{2}}).\label{kdv-d7}
\end{eqnarray}
In the above relations among $v_n$, $u_n$, $p_n$, $q_n$, $r_n$,
equations \eqref{kdv-d3}-\eqref{kdv-d4} are derived from
\eqref{kdv-d1}, whereas equations \eqref{kdv-d5}-\eqref{kdv-d7} come
from \eqref{kdv-d2}. Replacing $u_n$ by $u(x,t)$, $u_{n+h}$ by
$u(x+h,t)$, and similarly for the other variables, if we take the
limit $h\rightarrow 0$ in the above equations, we get
$u_t=\frac{1}{4}r+3up$, $v_t=\frac{1}{4}q+3u^2/2$, $p=u_x$, $q=p_x$,
$r=q_x$. Thus \eqref{kdv-d3}-\eqref{kdv-d7} construct a discretization of the KdV
equation \eqref{kdv11}-\eqref{kdv13}.
From the discussion above, we have the following theorem:
\begin{thm}
The system \eqref{kdv-d1}-\eqref{kdv-d2} is an integrable discretization of the KdV equation \eqref{kdv2}. Using
transformations  $v_n=(ln f_n)_{x}$, $ u_n=v_{n,x}$, $p_n=u_{n,x}$, $ q_n=p_{n,x}$,
this system can be converted to \eqref{kdv-d3}-\eqref{kdv-d7}, which converge to the KdV equation \eqref{kdv11}-\eqref{kdv13} as $h\rightarrow 0$.
\end{thm}

As an integrable system, we derived its bilinear B\"acklund transformation.
\begin{prop}
\label{kdv-prop} The bilinear equations \eqref{kdv-d1} and
\eqref{kdv-d2} have the B\"acklund transformation (BT)
\begin{eqnarray}
&& (D_x e^{-\frac{h}{2}D_n})f_n\cdot g_n=(-\frac{1}{h} e^{-\frac{h}{2}D_n}+\beta e^{\frac{h}{2}D_n})f_n\cdot g_n,\label{dkdv-bt1}\\
&&D_x^2f_n \cdot g_n =\gamma f_ng_n,\label{dkdv-bt2}\\
&& (D_t-\frac{1}{4}D_x^3-\frac{3\gamma}{4}D_x)f_n\cdot g_n=0,\label{dkdv-bt3}
\end{eqnarray}
where $\beta$ and $\gamma$  are arbitrary constants.
\end{prop}
{\bf Proof.} Let $f_n$ be a solution of equations
\eqref{kdv-d1}-\eqref{kdv-d2}. If we can show that the $g_n$ given
by \eqref{dkdv-bt1}-\eqref{dkdv-bt3} satisfies
\begin{eqnarray*}
&& P_1\equiv  D_x(D_t-\frac{1}{4}D_x^3)g_n \cdot g_n=0,\\
&& P_2\equiv (D_x^2-\frac{2}{h} D_x)e^{\frac{h}{2}D_n}g_n \cdot
g_{n}=0,
\end{eqnarray*}
then \eqref{dkdv-bt1}-\eqref{dkdv-bt3} form a BT for
\eqref{kdv-d1}-\eqref{kdv-d2}. It has been shown in \cite{d1} that
equations \eqref{dkdv-bt2}-\eqref{dkdv-bt3} construct a BT for
$D_x(D_t-\frac{1}{4}D_x^3)f \cdot f=0$. Thus $P_1=0$ is obvious. As
for $P_2$, by using \eqref{dkdv-bt1}-\eqref{dkdv-bt3} and the
bilinear identities \eqref{a1}-\eqref{a3}, we can precisely
calculate that
\begin{eqnarray*}
&\quad-[e^{\frac{D_n}{2}}f_n\cdot f_n]P_2 \\
&\equiv [(D_x^2-\frac{2}{h} D_x)e^{\frac{h}{2}D_n}f_n \cdot f_{n}][e^{\frac{D_n}{2}}g_n\cdot g_n]\nonumber\\
& \quad- [e^{\frac{D_n}{2}}f_n\cdot f_n][(D_x^2-\frac{2}{h} D_x)e^{\frac{h}{2}D_n}g_n \cdot g_{n}]\\
&=D_x[(D_xe^{\frac{h}{2}D_n}f_n\cdot g_n)\cdot(e^{-\frac{h}{2}D_n}f_n\cdot g_n)-(e^{\frac{h}{2}D_n}f_n\cdot g_n)\cdot(D_x^{-\frac{h}{2}D_n}f_n\cdot g_n)]\nonumber\\
& \quad -\frac{2}{h}D_x(e^{\frac{h}{2}D_n}f_n\cdot g_n)\cdot(e^{-\frac{h}{2}D_n}f_n\cdot g_n)\\
&=D_x[(D_xe^{\frac{h}{2}D_n}f_n\cdot g_n)\cdot(e^{-\frac{h}{2}D_n}f_n\cdot g_n)+(e^{\frac{h}{2}D_n}f_n\cdot g_n)\cdot(D_x^{-\frac{h}{2}D_n}f_n\cdot g_n)]\nonumber\\
& \quad -2D_x(e^{\frac{h}{2}D_n}f_n\cdot g_n)\cdot(D_xe^{-\frac{h}{2}D_n}f_n\cdot g_n+\frac{1}{h}e^{-\frac{h}{2}D_n}f_n\cdot g_n)\\
&=2\sinh(\frac{h}{2}D_n)[(D_x^2)f_n\cdot g_n]\cdot (f_ng_n)=0.
\end{eqnarray*}
Thus we have completed the proof of \textbf{Proposition 1}.
\begin{rem}
The way to derive B\"acklund transformations we used here is constructed by Hirota in \cite{hr5}.
\end{rem}

Setting $v_n=(ln g_n)_x$, $ u_n=v_{n,x}$, $p_n=u_{n,x}$, $
q_n=p_{n,x}$, $f_n=\phi_ng_n$ and $\psi_n=\phi_{n,x}$ in
\eqref{dkdv-bt1}-\eqref{dkdv-bt3}, we can get a Lax pair for the
discrete equations \eqref{kdv-d3}-\eqref{kdv-d7}:
\begin{displaymath}
\beta\left( \begin{array}{c}
\phi_{n+1}\\
\psi_{n+1}
\end{array} \right) =
\left(
\begin{array}{cc}
 \frac{1}{h}+v_n-v_{n+1} & \quad 1 \\
 \gamma -u_n-u_{n+1} & \quad \frac{1}{h}+v_n-v_{n+1} \\
\end{array}
\right)
\left( \begin{array}{c}
\phi_{n}\\
\psi_{n}
\end{array} \right),
\end{displaymath}
\begin{displaymath}
\left( \begin{array}{c}
\phi_{n,t}\\
\psi_{n,t}
\end{array} \right) =
\left(
\begin{array}{cc}
 -\frac{1}{2} p_n & \quad \gamma +u_n \\
 -\frac{1}{2} q_n+(\gamma -2 u_n) (\gamma +u_n) & \quad \frac{1}{2} p_n \\
\end{array}
\right)
\left( \begin{array}{c}
\phi_{n}\\
\psi_{n}
\end{array} \right).
\end{displaymath}
Equations \eqref{kdv-d3}-\eqref{kdv-d7} can be obtained from the compatibility of the two linear problems above.

We summarize the procedure in the following algorithm:
\begin{algorithm}
Given an integrable equation, let $x$ be the smooth variable, $n$ be the corresponding discrete variable and $h$ be the step size.\\
\textbf{Step 1}. Transform the integrable system to bilinear form via variable transformations.\\
\textbf{Step 2}. Derive the bilinear B\"acklund transformation and construct a new compatible system from the original equation and the B\"acklund transformation.\\
\textbf{Step 3}. Choose parameters for the B\"acklund transformation to satisfy the convergence condition
\begin{equation*}
D_x=D_n+O(h^k), \ \ k\geq1.
\end{equation*}
\textbf{Step 4}. Transform the new integrable system to nonlinear form and check whether the nonlinear form converges to the original equation as $h\rightarrow 0$.
\end{algorithm}
\begin{rem}
In the case of the KdV equation, only the first equation \eqref{kdv-b1} of the B\"acklund transformation is used to construct a new compatible system. However, in other cases like the Kadomtsev-Petviashvili equation (see Section 2), all member equations of its B\"acklund transformation may be needed to make the discrete system closed and one may even employ other compatible equations to construct the integrable discretization.
\end{rem}

We apply the newly proposed procedure to several equations. In the following 4 sections, we consider the extended Kadomtsev-Petviashvili (KP) equation, the extended Boussinesq equation, the extended Sawada-Kotera equation and the extended Ito equation, respectively.

\section{Kadomtsev-Petviashvili Equation}
\setcounter{equation}{0}
In this section, we attempt to use the new approach to derive a discretization of the extended KP equation
\begin{eqnarray}
&& 4u_t-r-12up-3v_{yy}=0,\label{ekp11}\\
&& p=u_x, q=p_x, r=q_x,u_y= v_{xy}.\label{ekp12}
\end{eqnarray}

The dependent variable transformation
\begin{equation}
u=(ln f)_{xx}
\end{equation}
gives the bilinear form
\begin{equation}
 (D_x^4-4D_x D_t+3D_y^2)f \cdot f=0.\label{kp11}
\end{equation}
We will call the system \eqref{ekp11}-\eqref{ekp12} and its bilinear form \eqref{kp11} KP equation when there is no ambiguity.
A bilinear B\"acklund transformation of \eqref{kp11} is
\begin{eqnarray}
&& (D_y-D_x^2-\mu D_x)f \cdot g=0,\label{kp-b1}\\
&& (3D_yD_x-4D_t+D_x^3+3\mu D_y)f \cdot g=0,\label{kp-b2}
\end{eqnarray}
where $\mu$  is an arbitrary constant \cite{kp}.

Setting $f\rightarrow f_{n}$ and $g\rightarrow f_{n-h}$ in \eqref{kp11}, \eqref{kp-b1} and \eqref{kp-b2}, we get a compatible system
\begin{eqnarray}
&& (D_x^4-4D_x D_t+3D_y^2)f_n \cdot f_n=0,\label{kp-c1}\\
&& (D_y-D_x^2-\mu D_x)f_{n} \cdot f_{n-h}=0,\label{kp-c2}\\
&& (3D_yD_x-4D_t+D_x^3+3\mu D_y)f_{n} \cdot f_{n-h}=0,\label{kp-c3}
\end{eqnarray}
where $h$ is the step size. Now rewrite \eqref{kp-c2} as
\begin{equation*}
[D_y\sinh(\frac{h}{2}D_n)-D_x^2\cosh(\frac{h}{2}D_n)-\mu D_x\sinh(\frac{h}{2}D_n)]f_{n} \cdot f_{n}=0.
\end{equation*}
Expanding this equation in powers of $h$, we have
\begin{equation}
[D_x^2+\frac{h}{2}\mu D_xD_n+O(h)]f_n\cdot f_n=0.
\end{equation}
By considering the convergence condition $D_x=D_n+O(h^k)$, $k\geq1$, we have $\mu=-\frac{2}{h}$. Thus we get an integrable differential-difference system
\begin{eqnarray}
&& (D_x^4-4D_x D_t+3D_y^2)f_n \cdot f_n=0,\label{kp-d1}\\
&& (D_y-D_x^2+\frac{2}{h} D_x)f_{n} \cdot f_{n-h}=0,\label{kp-d2}\\
&& (3D_yD_x-4D_t+D_x^3-\frac{6}{h}D_y)f_{n} \cdot f_{n-h}=0.\label{kp-d3}
\end{eqnarray}
We move on to show that this system (viewing $x$ as an auxiliary variable) converges to the KP equation in nonlinear form as $h\rightarrow 0$.
Letting $ w_n=ln(f_n)$, $v_n= w_{n,x}$, $ u_n=v_{n,x}$, $ p_n= u_{n,x}$, $ q_n= p_{n,x}$, $r_n=q_{n,x}$, equations
\eqref{kp-d1}-\eqref{kp-d3} can be transformed to
\begin{eqnarray}
&& 4u_{n,t}-r_n-12u_np_n-3v_{n,yy}=0,\label{kp-d4}\\
&& u_{n+h}+ u_{n}=\frac{2}{h}(v_{n+h}-v_{n})+( w_{n+h,y}- w_{n,y})
-(v_{n+h}-v_{n})^2,\label{kp-d5}\\
&& p_{n+h}+ p_{n}=\frac{2}{h}( u_{n+h}-u_{n})+(v_{n+h,y}-v_{n,y})\nonumber\\
&&\qquad \qquad \qquad\quad-2(v_{n+h}-v_{n})( u_{n+h}- u_{n}),\label{kp-d6}\\
&& q_{n+h}+ q_{n}=\frac{2}{h}( p_{n+h}-p_{n})+(u_{n+h,y}-u_{n,y})\nonumber\\
&&\qquad \qquad \qquad\quad-2(v_{n+h}-v_{n})( p_{n+h}- p_{n})-2( u_{n+h}- u_{n})^2,\label{kp-d7}\\
&& r_{n+h}+ r_{n}=\frac{2}{h}( q_{n+h}-q_{n})+(p_{n+h,y}-p_{n,y})\nonumber\\
&&\qquad \qquad \qquad\quad-2(v_{n+h}-v_{n})( q_{n+h}- q_{n})\nonumber\\
&&\qquad \qquad \qquad\quad-6( u_{n+h}- u_{n})(p_{n+h}-p_{n}),\label{kp-d8}\\
&&3((v_{n+h,y}+v_{n,y})+( w_{n+h,y}- w_{n,y})
(v_{n+h}-v_{n}))-4( w_{n+h,t}- w_{n,t})\nonumber\\
&&\quad +( p_{n+h}- p_{n})+3(v_{n+h}-v_{n})( u_{n+h}+ u_{n})+(v_{n+h}-v_{n})^3\nonumber\\
&&\quad =\frac{6}{h}( w_{n+h,y}- w_{n,y}).\label{kp-d9}
\end{eqnarray}
Deriving an analog of \eqref{kp-d9} with $x$ instead and eliminating $w$ by using \eqref{kp-d5}, we get
\begin{eqnarray}
&&3(u_{n+h,y}+u_{n,y})+6( u_{n+h}+ u_{n})
(u_{n+h}-u_{n})+6( u_{n+h}- u_{n})
(v_{n+h}-v_{n})^2\nonumber\\
&& \quad +3( v_{n+h,y}- v_{n,y})
(v_{n+h}-v_{n})-4( v_{n+h,t}- v_{n,t})+( q_{n+h}- q_{n})\nonumber\\
&&\quad +3(v_{n+h}-v_{n})( p_{n+h}+ p_{n})-\frac{6}{h}( u_{n+h}- u_{n})(v_{n+h}-v_{n})\nonumber\\
&&\quad =\frac{6}{h}( v_{n+h,y}- v_{n,y}).\label{kp-d10}
\end{eqnarray}

Replacing $v_n$ by $v(x,y,t)$, $v_{n+h}$ by $v(x+h,y,t)$, and similarly for $u$, $p$, $q$, $r$ and then taking the limit $h\rightarrow0$, equations \eqref{kp-d4}, \eqref{kp-d6}-\eqref{kp-d8} and \eqref{kp-d10} become
\begin{eqnarray}
&& 4u_t-r-12up-3v_{yy}=0,\label{kp-e1}\\
&& p=u_x, q=p_x, r=q_x, u_y= v_{xy}.\label{kp-e2}
\end{eqnarray}

From the discussion above, we have the following theorem for the KP equation:
\begin{thm}
The system \eqref{kp-d1}-\eqref{kp-d3} is an integrable discretization of the KP equation \eqref{kp11}. Using
transformations $v_n= w_{n,x}$, $ u_n=v_{n,x}$, $ p_n= u_{n,x}$, $ q_n= p_{n,x}$, $r_n=q_{n,x}$,
this system can be converted to \eqref{kp-d4}, \eqref{kp-d6}-\eqref{kp-d8} and \eqref{kp-d10}, which converge to the KP equation \eqref{ekp11}-\eqref{ekp12} as $h\rightarrow 0$.
\end{thm}
\begin{rem}
Unlike the KdV equation case, we need both equations of the B\"acklund transformation to construct the integrable discretization of the KP equation. Note that there are 5 variables $v_n$, $u_n$, $p_n$, $q_n$, $r_n$ in \eqref{kp-d4}, \eqref{kp-d6}-\eqref{kp-d8} and \eqref{kp-d10}. The purpose of \eqref{kp-d10} is to make the system closed.
\end{rem}

We have the following proposition for \eqref{kp-d1}-\eqref{kp-d3}:
\begin{prop}
Bilinear equations \eqref{kp-d1}-\eqref{kp-d3} have the B\"acklund
transformation
\begin{eqnarray}
&& (D_x e^{-\frac{h}{2}D_n})f_n\cdot g_n=(-\frac{1}{h} e^{-\frac{h}{2}D_n}+\beta e^{\frac{h}{2}D_n})f_n\cdot g_n,\label{dkp-bt1}\\
&&(D_y-D_x^2)f_n \cdot g_n =\gamma f_ng_n,\label{dkp-bt2}\\
&& (3D_yD_x-4D_t+D_x^3-3\gamma D_x)f_n\cdot g_n=0,\label{dkp-bt3}
\end{eqnarray}
where $\beta$ and $\gamma$  are arbitrary constants.
\end{prop}
{\bf Proof.} Let $f_n$ be a solution of
\eqref{kp-d1}-\eqref{kp-d2} and $g_n$ be given by
\eqref{dkp-bt1}-\eqref{dkp-bt3}.
In a way similar to the proof of Proposition 1, we have
\begin{eqnarray*}
&& P_1\equiv (D_x^4-4D_x D_t+3D_y^2)g_n \cdot g_n=0,\\
&& P_2\equiv (D_y-D_x^2+\frac{2}{h} D_x)e^{\frac{h}{2}D_n}g_n \cdot g_{n}=0.
\end{eqnarray*}
Thus it suffices to show that
\begin{eqnarray*}
&& P_3\equiv (3D_yD_x-4D_t+D_x^3-\frac{6}{h}D_y)e^{\frac{h}{2}D_n}g_n \cdot g_{n}=0.
\end{eqnarray*}
In fact, by using $P_2=0$, equations \eqref{dkp-bt1}-\eqref{dkp-bt3} and the bilinear
identities \eqref{a1}-\eqref{a7}, we can precisely deduce
\begin{eqnarray*}
&&\quad-[e^{\frac{D_n}{2}}f_n\cdot f_n]P_3 \\
&&\equiv[(3D_yD_x-4D_t+D_x^3-\frac{6}{h}D_y)e^{\frac{h}{2}D_n}f_n \cdot f_{n}][e^{\frac{D_n}{2}}g_n\cdot g_n]\nonumber\\
&&\quad- [e^{\frac{D_n}{2}}f_n\cdot f_n][(3D_yD_x-4D_t+D_x^3-\frac{6}{h}D_y)e^{\frac{h}{2}D_n}g_n \cdot g_{n}]\nonumber\\
&&=2\sinh(\frac{h}{2}D_n)[(3D_xD_y-4D_t+D_x^3-3\gamma D_x)f_n\cdot g_n]\cdot(f_ng_n)=0.
\end{eqnarray*}
Hence the proof is complete.

Setting $v_n=(ln g_n)_x$, $ u_n=v_{n,x}$, $p_n=u_{n,x}$, $
q_n=p_{n,x}$, $f_n=\phi_ng_n$ and $\psi_n=\phi_{n,x}$ in
\eqref{dkp-bt1}-\eqref{dkp-bt3}, we can get a Lax pair for the
discrete equations \eqref{kp-d4}, \eqref{kp-d6}-\eqref{kp-d8} and
\eqref{kp-d10}, namely,
\begin{displaymath}
\beta\left( \begin{array}{c}
\phi_{n+1}\\
\psi_{n+1}
\end{array} \right) =
\left(
\begin{array}{cc}
 \frac{1}{h}+v_n-v_{n+1} & \quad 1 \\
 \partial_y-u_n-u_{n+1}-\gamma &\quad \frac{1}{h}+v_n-v_{n+1} \\
\end{array}
\right)
\left( \begin{array}{c}
\phi_{n}\\
\psi_{n}
\end{array} \right)
\end{displaymath}
and
\begin{displaymath}
\left( \begin{array}{c}
\phi_{n,t}\\
\psi_{n,t}
\end{array} \right) =
\left(
\begin{array}{cc}
 \frac{3}{2}v_{n,y}-\frac{1}{2} p_n & \quad \partial_y-\gamma +u_n \\
 M_{21} & \quad\frac{3}{2}v_{n,y}+\frac{1}{2} p_n \\
\end{array}
\right)
\left( \begin{array}{c}
\phi_{n}\\
\psi_{n}
\end{array} \right).
\end{displaymath}
where $M_{21}=-\frac{1}{2}u_{n,y}-\frac{1}{2} q_n+(\gamma -u_n) (\gamma +2u_n)-(2\gamma +u_n)\partial_y+\partial_y^2$.

\section{Boussinesq equation}
\setcounter{equation}{0}
The extended Boussinesq equation is
\begin{eqnarray}
&& u_{tt}-q-s-12uq-12p^2=0,\label{ebs-1}\\
&& v_{tt}-p-r-12up=0,\label{ebs-2}\\
&& p=u_x, q=p_x, r=q_x, s=r_x.\label{ebs-3}
\end{eqnarray}

Under the dependent variable transformation
\begin{equation}
u=(ln f)_{xx} \ ,v=(ln f)_{x} \ ,
\end{equation}
equation \eqref{ebs-1}-\eqref{ebs-3} can be transformed to the bilinear form
\begin{equation}
 (D_t^2-D_x^2-D_x^4)f \cdot f=0,\label{bs11}
\end{equation}
We will still call the extended Boussinesq equation and its bilinear form  Boussinesq equation without confusion. A two-parameter  B\"acklund transformation has been given in \cite{bs1}:
\begin{eqnarray}
&& (D_t-a D_x^2+\xi D_x)f\cdot g=0,\label{bs-b1}\\
&& (-aD_xD_t+D_x^3+\xi a D_x^2+(1-\xi^2)D_x-\eta a)f\cdot g=0,\label{bs-b2}
\end{eqnarray}
where $a^2 =-3$ and $\xi$, $\eta$ are arbitrary parameters.
We can rewrite \eqref{bs-b2} as
\begin{equation}
(3D_xD_t+aD_x^3+a D_x+a\xi D_t+3\eta)f\cdot g=0.\label{bs-b3}
\end{equation}
Setting $\eta=0$, $\xi=\frac{2a}{h}$, $f\rightarrow f_n$, $g\rightarrow f_{n-h}$ in \eqref{bs11}, \eqref{bs-b1}, we get a compatible differential-difference system
\begin{eqnarray}
&& (D_t^2-D_x^2-D_x^4)f_n \cdot f_n=0,\label{bs-d1}\\
&& (D_t-a D_x^2+\frac{2a}{h} D_x)f_n\cdot f_{n-h}=0.\label{bs-d2}
\end{eqnarray}
Here $\eta=0$, $\xi=\frac{2a}{h}$ are obtained from the convergence condition $D_x=D_n+O(h^k)$.

Viewing $x$ as an auxiliary variable, substituting $ w_n=2ln(f_n)$, $v_n= w_{n,x}$, $ u_n=v_{n,x}$, $ p_n= u_{n,x}$, $ q_n= p_{n,x}$, $r_n=q_{n,x}$
into \eqref{bs-d1}-\eqref{bs-d2}, we can get
\begin{eqnarray}
&& u_{n,tt}-q_n-s_n-12u_nq_n-12p_n^2=0,\label{bs-d4}\\
&& v_{n,tt}-p_n-r_n-12u_np_n=0,\label{bs-d5}\\
&& a(p_{n+h}+ p_{n})
=\frac{2a}{h}(u_{n+h}-u_{n})+(v_{n+h,t}-v_{n,t})\nonumber\\
&&\qquad\qquad\qquad\qquad\quad-2a(u_{n+h}-u_{n})(v_{n+h}-v_{n}),\label{bs-d6}\\
&& a(q_{n+h}+q_{n})
=\frac{2a}{h}(p_{n+h}-p_{n})+(u_{n+h,t}-u_{n,t})\nonumber\\
&&\qquad\qquad\qquad\qquad\quad -2a(p_{n+h}-p_{n})(v_{n+h}-v_{n})-2a(u_{n+h}-u_{n})^2,\label{bs-d7}\\
&& a(r_{n+h}+r_{n})
=\frac{2a}{h}(q_{n+h}-q_{n})+(p_{n+h,t}-p_{n,t})\nonumber\\
&&\qquad\qquad\qquad\qquad\quad
-6a(p_{n+h}-p_{n})(u_{n+h}-u_{n})\nonumber\\
&&\qquad\qquad\qquad\qquad\quad
-2a(q_{n+h}-q_{n})(v_{n+h}-v_{n}),\label{bs-d8}\\
&& a(s_{n+h}+s_{n})
=\frac{2a}{h}(r_{n+h}-r_{n})+(q_{n+h,t}-q_{n,t})
-6a(p_{n+h}-p_{n})^2\nonumber\\
&&\qquad\qquad\qquad\qquad\quad
-8a(q_{n+h}-q_{n})(u_{n+h}-u_{n})\nonumber\\
&&\qquad\qquad\qquad\qquad\quad
-2a(r_{n+h}-r_{n})(v_{n+h}-v_{n}).\label{bs-d9}
\end{eqnarray}
Here \eqref{bs-d6}-\eqref{bs-d9} are all derived from \eqref{bs-d2}.

Replacing $u_n$ by $u(x,t)$, $u_{n+h}$ by $u(x+h,t)$, and similarly for $w$, $v$, $p$, $q$, $r$, $s$ and then taking the limit $h\rightarrow0$, equations \eqref{bs-d4}-\eqref{bs-d9} become
\begin{eqnarray}
&& u_{tt}-q-s-12uq-12p^2=0,\label{bs-e1}\\
&& v_{tt}-p-r-12up=0,\label{bs-e2}\\
&& p=u_x, q=p_x, r=q_x, s=r_x.\label{bs-e3}
\end{eqnarray}

From the above discussion, we have the following theorem for the Boussinesq equation:
\begin{thm}
The system \eqref{bs-d1}-\eqref{bs-d2} is an integrable discretization of the Boussinesq equation \eqref{bs11}. Using
transformations $ w_n=2ln(f_n)$, $v_n= w_{n,x}$, $ u_n=v_{n,x}$, $ p_n= u_{n,x}$, $ q_n= p_{n,x}$, $r_n=q_{n,x}$,
this system can be transformed to \eqref{bs-d4}-\eqref{bs-d9}, which converge to the extended Boussinesq equation \eqref{ebs-1}-\eqref{ebs-3} as $h\rightarrow 0$.
\end{thm}

For the system \eqref{bs-d1}-\eqref{bs-d2}, we have the following proposition:
\begin{prop}
The bilinear equations \eqref{bs-d1}-\eqref{bs-d2} have the
B\"acklund transformation
\begin{eqnarray}
&& (D_t-a D_x^2)f_n\cdot g_n=\lambda f_ng_n,\label{dbs-bt1}\\
&& (aD_xD_t-D_x^3-(\lambda a+1)D_x)f_n\cdot g_n=0,\label{dbs-bt2}\\
&& (D_x e^{-\frac{h}{2}D_n})f_n\cdot g_n=(-\frac{1}{h} e^{-\frac{h}{2}D_n}+\beta e^{\frac{h}{2}D_n})f_n\cdot g_n,\label{dbs-bt3}
\end{eqnarray}
where $\beta$ and $\lambda$  are arbitrary constants.
\end{prop}
{\bf Proof.} Let $f_n$ be a solution of \eqref{bs-d1}-\eqref{bs-d2}
and $g_n$ be given by \eqref{dbs-bt1}-\eqref{dbs-bt3}. Just like the
proof of \textbf{Proposition 1}, we can have
\begin{eqnarray*}
&& P_1\equiv (D_t^2-D_x^2-D_x^4)g_n \cdot g_n=0,\\
&& P_2\equiv (D_t-a D_x^2+\frac{2a}{h} D_x)e^{\frac{h}{2}D_n}g_n \cdot g_{n}=0,
\end{eqnarray*}
which means that equations \eqref{dbs-bt1}-\eqref{dbs-bt3} construct
a B\"acklund transformation of \eqref{bs-d1}-\eqref{bs-d2}.

We can derive a Lax pair for the nonlinear equations
\eqref{bs-d4}-\eqref{bs-d9} from the  B\"acklund transformation
\eqref{dbs-bt1}-\eqref{dbs-bt3}. Setting $v_n=(ln g_n)_x$, $
u_n=v_{n,x}$, $p_n=u_{n,x}$, $q_n=p_{n,x}$, $f_n=\phi_ng_n$,
$\psi_n=\phi_{n,x}$ and $\Delta_n=\psi_{n,x}$ in
\eqref{dbs-bt1}-\eqref{dbs-bt3}, we get
\begin{displaymath}
\beta\left( \begin{array}{c}
\phi_{n+1}\\
\psi_{n+1}\\
\Delta_{n+1}
\end{array} \right) =
L_n
\left( \begin{array}{c}
\phi_{n}\\
\psi_{n}\\
\Delta_n
\end{array} \right)
\end{displaymath}
and
\begin{displaymath}
\left( \begin{array}{c}
\phi_{n,t}\\
\psi_{n,t}\\
\Delta_{n,t}
\end{array} \right) =
Q_n
\left( \begin{array}{c}
\phi_{n}\\
\psi_{n}\\
\Delta_n
\end{array} \right),
\end{displaymath}
where
\begin{displaymath}
L_n =
\left(
\begin{array}{ccc}
 \frac{1}{h}+v_n-v_{n+1} & \quad 1 & \quad 0\\
 u_n-u_{n+1} & \quad \frac{1}{h}+v_n-v_{n+1} & \quad 1 \\
 -\frac{1}{2}p_n-p_{n+1}+\frac{a}{2}v_{n,t} & \quad -\frac{1}{4}-u_n-2u_{n+1} & \quad \frac{1}{h}+v_n-v_{n+1}
\end{array}
\right)
\end{displaymath}
\begin{displaymath}
Q_n =
\left(
\begin{array}{ccc}
 \lambda+2au_n & \quad 0&\quad a \\
 \frac{a}{2}p_n-\frac{3}{2}v_{n,t}  & \quad \lambda-au_n-\frac{a}{4} & \quad 0 \\
 \frac{a}{2}q_n-\frac{3}{2}u_{n,t}  & \quad -\frac{a}{2}p_n-\frac{3}{2}v_{n,t}  & \quad \lambda-au_n-\frac{a}{4}
\end{array}
\right).
\end{displaymath}

The compatibility condition of the two linear problems above is
$L_{n,t}=Q_{n+1}L_n-L_nQ_n$, which produces the discrete equations
\eqref{bs-d4}-\eqref{bs-d9}.

\section{Sawada-Kotera Equation}
\setcounter{equation}{0} In this section we apply the new method to
the extended Sawada-Kotera (SK) equation
\begin{eqnarray}
&& v_{t}+s+30uq+60u^3=0,\label{sk-e1}\\
&& u_{t}+\eta+30pq+30ur+180u^2p=0,\label{sk-e2}\\
&& u=v_x, p=u_x, q=p_x, r=q_x, s=r_x.\label{sk-e3}
\end{eqnarray}
The dependent variable transformation
\begin{equation}
u=(ln f)_{xx}, v=(ln f)_{x},
\end{equation}
gives the bilinear form
\begin{equation}
 D_x(D_t+D_x^5)f \cdot f=0.\label{sk11}
\end{equation}
A bilinear B\"acklund transformation of the extended SK equation is
\begin{eqnarray}
&& (D_x^3-\sigma D_x^2+\frac{1}{3}\sigma^2D_x+\lambda)f \cdot g=0,\label{sk-b1}\\
&& (2D_t-3D_x^5+5\sigma D_x^4-\frac{5}{3}\sigma^2D_x^3+15\lambda D_x^2-10\lambda \sigma D_x+\mu)f \cdot g=0,\label{sk-b2}
\end{eqnarray}
where $\kappa$, $\lambda$, $\sigma$ and $\mu$ are arbitrary
constants (see details in \cite{h3}). Setting $\lambda=0$, $\mu=0$,
$\sigma=-3\kappa$, $f\rightarrow f_{n}$ and $g\rightarrow f_{n-h}$
in \eqref{sk11}, \eqref{sk-b1} and \eqref{sk-b2}, we arrive at the
compatible system
\begin{eqnarray}
&&  D_x(D_t+D_x^5)f_n \cdot f_n=0,\label{sk-c1}\\
&& (D_x^3+3\kappa D_x^2+3\kappa^2D_x)f_n \cdot f_{n-h}=0,\label{sk-c2}\\
&& (2D_t-3D_x^5-15\kappa D_x^4-15\kappa^2D_x^3)f_n \cdot f_{n-h}=0.\label{sk-c3}
\end{eqnarray}
Now rewrite \eqref{sk-c2} as
\begin{equation*}
[D_x^3\sinh(\frac{h}{2}D_n)+3\kappa
D_x^2\cosh(\frac{h}{2}D_n)+3\kappa^2 D_x\sinh(\frac{h}{2}D_n)]f_{n}
\cdot f_{n}=0.
\end{equation*}
Expanding this equation in powers of $h$, we obtain
\begin{equation}
[3\kappa D_x^2+\frac{3h}{2}\kappa^2 D_xD_n+O(h)]f_n\cdot f_n=0.
\end{equation}
By imposing the convergence condition $D_x=D_n+O(h^k)$, $k\geq1$, we
find that $\kappa=-\frac{2}{h}$. Thus we get an integrable
differential-difference system
\begin{eqnarray}
&&  D_x(D_t+D_x^5)f_n \cdot f_n=0,\label{sk-d1}\\
&& (D_x^3-\frac{6}{h} D_x^2+\frac{12}{h^2}D_x)f_{n} \cdot f_{n-h}=0,\label{sk-d2}\\
&& (2D_t-3D_x^5+\frac{30}{h}D_x^4-\frac{60}{h^2}D_x^3)f_{n} \cdot f_{n-h}=0.\label{sk-d3}
\end{eqnarray}
Letting $v_n= (ln f_n)_x$, $u_n=v_{n,x}$, $p_n= u_{n,x}$, $q_n=
p_{n,x}$, $r_n= q_{n,x}$, $s_n= r_{n,x}$, $\eta_n= s_{n,x}$,
equations \eqref{sk-d1}-\eqref{sk-d2} can be transformed to

\begin{eqnarray}
&& v_{n,t}+s_n+30u_nq_n+60u_n^3=0,\label{sk-d4}\\
&& u_{n,t}+\eta_n+30p_nq_n+30u_nr_n+180u_n^2p_n=0,\label{sk-d5}\\
&&(p_{n+h}-p_{n})+3(v_{n+h}-v_{n})(u_{n+h}
+u_{n})+(v_{n+h}-v_{n})^3\nonumber\\
&&\quad-\frac{6}{h}((u_{n+h}+u_{n})+(v_{n+h}-v_{n})^2)
+\frac{12}{h^2}(v_{n+h}-v_{n})=0,\label{sk-d6}\\
&&(q_{n+h}-q_{n})
+3(u_{n+h}-u_{n})(u_{n+h}+u_{n})
\nonumber\\
&&\quad+3(v_{n+h}-v_{n})(p_{n+h}+p_{n})
+3(v_{n+h}-v_{n})^2(u_{n+h}-u_{n})\nonumber\\
&&\quad-\frac{6}{h}((p_{n+h}+p_{n})
+2(v_{n+h}-v_{n})(u_{n+h}-u_{n}))\nonumber\\
&&\quad +\frac{12}{h^2}(u_{n+h}-u_{n})=0,\label{sk-d7}\\
&&(r_{n+h}-r_{n})
+3(p_{n+h}-p_{n})(u_{n+h}+u_{n})
+3(v_{n+h}-v_{n})(q_{n+h}+q_{n})\nonumber\\
&&\quad+6(u_{n+h}-u_{n})(p_{n+h}+p_{n})
+6(v_{n+h}-v_{n})(u_{n+h}-u_{n})^2\nonumber\\
&&\quad+3(v_{n+h}-v_{n})^2(p_{n+h}-p_{n})\nonumber\\
&&\quad-\frac{6}{h}((q_{n+h}+q_{n})
+2(v_{n+h}-v_{n})(p_{n+h}-p_{n})
+2(u_{n+h}-u_{n})^2)\nonumber\\
&&\quad +\frac{12}{h^2}(p_{n+h}-p_{n})=0,\label{sk-d8}
\end{eqnarray}
\begin{eqnarray}
&&(s_{n+h}-s_{n})
+3(q_{n+h}-q_{n})(u_{n+h}+u_{n})
+9(p_{n+h}-p_{n})(p_{n+h}+p_{n})\nonumber\\
&&\quad +9(u_{n+h}-u_{n})(q_{n+h}+q_{n})
+3(v_{n+h}-v_{n})(r_{n+h}+r_{n})\nonumber\\
&&\quad +6(u_{n+h}-u_{n})^3
+18(v_{n+h}-v_{n})(u_{n+h}-u_{n})(p_{n+h}-p_{n})\nonumber\\
&&\quad
+3(v_{n+h}-v_{n})^2(q_{n+h}-q_{n})\nonumber\\
&&\quad-\frac{6}{h}((r_{n+h}+r_{n})
+2(v_{n+h}-v_{n})(q_{n+h}-q_{n})
+6(u_{n+h}-u_{n})(p_{n+h}-p_{n}))\nonumber\\
&&\quad
+\frac{12}{h^2}(q_{n+h}-q_{n})=0,\label{sk-d9}\\
&&(\eta_{n+h}-\eta_{n})
+3(r_{n+h}-r_{n})(u_{n+h}+u_{n})
+12(q_{n+h}-q_{n})(p_{n+h}+p_{n})\nonumber\\
&&\quad +18(p_{n+h}-p_{n})(q_{n+h}+q_{n})
+12(u_{n+h}-u_{n})(r_{n+h}+r_{n})\nonumber\\
&&\quad +3(v_{n+h}-v_{n})(s_{n+h}+s_{n})
+36(u_{n+h}-u_{n})^2(p_{n+h}-p_{n})\nonumber\\
&&\quad 18(v_{n+h}-v_{n})(p_{n+h}-p_{n})^2
+24(v_{n+h}-v_{n})(u_{n+h}-u_{n})(q_{n+h}-q_{n})
\nonumber\\
&&\quad+3(v_{n+h}-v_{n})^2(r_{n+h}-r_{n})
-\frac{6}{h}((s_{n+h}+s_{n})
+2(v_{n+h}-v_{n})(r_{n+h}-r_{n})\nonumber\\
&&\quad+8(u_{n+h}-u_{n})(q_{n+h}-q_{n})
+6(p_{n+h}-p_{n})^2)+\frac{12}{h^2}(r_{n+h}-r_{n})=0.\label{sk-d10}
\end{eqnarray}
Replacing $v_n$ by $v(x,t)$, $v_{n+h}$ by $v(x+h,t)$, and similarly
for $u$, $p$, $q$, $r$, $s$, as $h\rightarrow0$, equations
\eqref{sk-d4}-\eqref{sk-d10} become
\begin{eqnarray}
&& v_{t}+s+30uq+60u^3=0,\\
&& u_{t}+\eta+30pq+30ur+180u^2p=0,\\
&& u=v_x, p=u_x, q=p_x, r=q_x, s=r_x.
\end{eqnarray}
\begin{rem}
For \eqref{sk-d6}-\eqref{sk-d10}, both sides must be multiplied by $h$
before taking the limit.
\end{rem}

For the bilinear differential-difference equations
\eqref{sk-d1}-\eqref{sk-d3}, we have this proposition:
\begin{prop}
The bilinear equations \eqref{sk-d1}-\eqref{sk-d3} have the
B\"acklund transformation
\begin{eqnarray}
&& D_x^3f_n\cdot g_n=\lambda f_ng_n,\label{dsk-bt1}\\
&& (2D_t-3D_x^5-15\lambda D_x^2)f_n\cdot g_n=0,\label{dsk-bt2}\\
&& (D_x e^{\frac{h}{2}D_n}-D_x e^{-\frac{h}{2}D_n})f_n\cdot g_n=\frac{2}{h} (e^{\frac{h}{2}D_n}+e^{-\frac{h}{2}D_n})f_n\cdot g_n,\label{dsk-bt3}\\
&& (D_x^3 e^{\frac{h}{2}D_n}-D_x^3
e^{-\frac{h}{2}D_n}-\frac{6}{h}D_x^2e^{\frac{h}{2}D_n}-\frac{6}{h}D_x^2e^{-\frac{h}{2}D_n}\nonumber\\
&&\qquad \qquad +2\lambda e^{\frac{h}{2}D_n}-2\lambda e^{-\frac{h}{2}D_n})f_n\cdot g_n=0,\label{dsk-bt4}
\end{eqnarray}
where $\lambda$ is an arbitrary constant.
\end{prop}
{\bf Proof.} Let $f_n$ be a solution of \eqref{sk-d1}-\eqref{sk-d3}
and $g_n$ be given by \eqref{dsk-bt1}-\eqref{dsk-bt4}. If we can
show that
\begin{eqnarray*}
&& P_1\equiv  D_x(D_t+D_x^5)g_n \cdot g_n=0,\\
&& P_2\equiv (D_x^3-\frac{6}{h} D_x^2+\frac{12}{h^2}D_x)e^{\frac{h}{2}D_n}g_n \cdot g_{n}=0,\\
&& P_3\equiv
(2D_t-3D_x^5+\frac{30}{h}D_x^4-\frac{60}{h^2}D_x^3)e^{\frac{h}{2}D_n}g_n
\cdot g_{n}=0,
\end{eqnarray*}
then equations \eqref{dsk-bt1}-\eqref{dsk-bt4} construct a
B\"acklund transformation of \eqref{sk-d1}-\eqref{sk-d3}. It is
obvious that $P_1=0$ as \eqref{dsk-bt1}-\eqref{dsk-bt2} simply form
a B\"acklund transformation of $D_x(D_t+D_x^5)f_n \cdot f_n=0$ (see
\eqref{sk-b1} and \eqref{sk-b2}).

By \eqref{dsk-bt1}-\eqref{dsk-bt4} and the bilinear identities
\eqref{a1}, \eqref{a2}, \eqref{a6}, we can precisely demonstrate
that
\begin{eqnarray*}
&&\quad -(e^{\frac{h}{2}D_n}f_n\cdot f_n)P_2\\
&&\equiv [(D_x^3-\frac{6}{h} D_x^2+\frac{12}{h^2}D_x)e^{\frac{h}{2}D_n}f_n \cdot f_n](e^{\frac{h}{2}D_n}g_n\cdot g_n)\\
&&\quad-(e^{\frac{h}{2}D_n}f_n\cdot f_n)[(D_x^3-\frac{6}{h} D_x^2+\frac{12}{h^2}D_x)e^{\frac{h}{2}D_n}g_n \cdot g_n]\\
&&=-3D_x[(D_xe^{\frac{h}{2}D_n}f_n\cdot
g_n-\frac{2}{h}e^{\frac{h}{2}D_n}f_n\cdot
g_n)\cdot(D_xe^{-\frac{h}{2}D_n}f_n\cdot
g_n+\frac{2}{h}e^{-\frac{h}{2}D_n}f_n\cdot g_n)]\nonumber\\
&&=0.
\end{eqnarray*}
By using \eqref{sk-d2}, \eqref{dsk-bt1}-\eqref{dsk-bt4} and the
bilinear identities \eqref{a1}, \eqref{a3}, \eqref{a6}, \eqref{a7},
\eqref{a11}, \eqref{a12}
\begin{eqnarray*}
&&\quad -(e^{\frac{h}{2}D_n}f_n\cdot f_n)P_3\\
&&\equiv [(2D_t-3D_x^5+\frac{30}{h}D_x^4-\frac{60}{h^2}D_x^3)e^{\frac{h}{2}D_n}f_n \cdot f_n)](e^{\frac{h}{2}D_n}g_n\cdot g_n)\\
&&\quad -(e^{\frac{h}{2}D_n}f_n\cdot f_n)[(2D_t-3D_x^5+\frac{30}{h}D_x^4-\frac{60}{h^2}D_x^3)e^{\frac{h}{2}D_n}g_n \cdot g_n]\\
&&=15D_x[(D_x^3e^{\frac{h}{2}D_n}-D_x^3e^{-\frac{h}{2}D_n}-\frac{6}{h}D_x^2e^{\frac{h}{2}D_n}-\frac{6}{h}D_x^2e^{-\frac{h}{2}D_n}+2\lambda e^{\frac{h}{2}D_n}\nonumber\\
&&\quad-2\lambda e^{-\frac{h}{2}D_n})f_n\cdot g_n]\cdot(D_xe^{-\frac{h}{2}D_n}f_n\cdot
g_n+\frac{2}{h}e^{-\frac{h}{2}D_n}f_n\cdot g_n)=0.
\end{eqnarray*}
Thus we have completed the proof. Here we omitted the detailed calculation.

From the  B\"acklund transformation \eqref{dsk-bt1}-\eqref{dsk-bt2},
we can derive a Lax pair for the system
\eqref{sk-d4}-\eqref{sk-d10}. Setting $v_n=(ln g_n)_x$, $
u_n=v_{n,x}$, $p_n=u_{n,x}$, $q_n=p_{n,x}$, $r_n=q_{n,x}$,
$s_n=r_{n,x}$, $\eta_n=s_{n,x}$, $f_n=\phi_ng_n$ ,
$\psi_n=\phi_{n,x}$, $\Delta_n=\psi_{n,x}$ in
\eqref{dsk-bt1}-\eqref{dsk-bt4}, we have
\begin{displaymath}
L_{1,n}
\left( \begin{array}{c}
\phi_{n+1}\\
\psi_{n+1}\\
\Delta_{n+1}
\end{array} \right)=L_{2,n}
\left( \begin{array}{c}
\phi_{n}\\
\psi_{n}\\
\Delta_n
\end{array} \right)
\end{displaymath}
and
\begin{displaymath}
\left( \begin{array}{c}
\phi_{n,t}\\
\psi_{n,t}\\
\Delta_{n,t}
\end{array} \right) =
Q_n
\left( \begin{array}{c}
\phi_{n}\\
\psi_{n}\\
\Delta_n
\end{array} \right),
\end{displaymath}
where
\begin{displaymath}
L_{1,n}=
\left(
\begin{array}{ccc}
 \frac{-2-h v_n+h v_{n+1}}{h} & \quad 1 & \quad 0 \\
 \frac{-h u_{n}+h u_{n+1}}{h} & \quad \frac{-2-h v_n+h v_{n+1}}{h} & \quad 1 \\
  a & \quad b & \quad \frac{2 (2+h v_n-h v_{n+1})}{h} \\
\end{array}
\right),
\end{displaymath}
\begin{displaymath}
L_{2,n}=
\left(
\begin{array}{ccc}
 -\frac{-2-h v_n+h v_{n+1}}{h} & \quad 1 & \quad 0 \\
 -\frac{-h u_{n}+h u_{n+1}}{h} & \quad -\frac{-2-h v_n+h v_{n+1}}{h} & \quad 1 \\
 c& \quad d & \quad -\frac{4+2 h v_n-2 h v_{n+1}}{h} \\
\end{array}
\right),
\end{displaymath}
\begin{displaymath}
Q_n =
\left(
\begin{array}{ccc}
 36 \lambda  u_{n} &  6 \left(q_{n}-6 u_{n}^2\right) &  9 (\lambda -2 p_{n}) \\
 9 \lambda  (\lambda +2 p_{n}) &  6 (r_{n}-3 (\lambda -2 p_{n}) u_{n}) &  -12 \left(q_{n}+3 u_{n}^2\right) \\
 6 \lambda  \left(q_{n}-6 u_{n}^2\right) &  Q_n(3,2) &  -6 (r_{n}+3 (\lambda +2 p_{n}) u_{n}) \\
\end{array}
\right)
\end{displaymath}
with
\begin{eqnarray}
&& Q_n(3,2)=3 \left(3 \lambda ^2+12 p_{n}^2+2 s_{n}+36 q_{n} u_{n}+72 u_{n}^3\right),\nonumber\\
&& a=\frac{-2 h \lambda +3 u_{n} (2+h v_n-h v_{n+1})+3 u_{n+1} (2+h v_n-h v_{n+1})}{h}\nonumber\\
&& \qquad +\frac{(v_n-v_{n+1})^2 (6+h v_n-h v_{n+1})}{h},\nonumber\\
&& b=\frac{-5 h u_{n}-h u_{n+1}-12 v_n-3 h v_n^2+12 v_{n+1}+6 h v_n v_{n+1}-3 h v_{n+1}^2}{h},\nonumber\\
&& c=-\frac{2 h \lambda +3 u_{n} (2+h v_n-h v_{n+1})+3 u_{n+1} (2+h v_n-h v_{n+1})}{h}\nonumber\\
&& \qquad -\frac{(v_n-v_{n+1})^2 (6+h v_n-h v_{n+1})}{h},\nonumber\\
&& d= -\frac{h u_{n}+5 h u_{n+1}+12 v_n+3 h v_n^2-12 v_{n+1}-6 h v_n
v_{n+1}+3 h v_{n+1}^2}{h} .\nonumber
\end{eqnarray}
The compatibility condition of the two linear problems above is
$$L_{2,n,t}L_{2,n}^{-1}-L_{1,n,t}L_{1,n}^{-1}=L_{1,n}Q_{n+1}L_{1,n}^{-1}-L_{2,n}Q_{n}L_{2,n}^{-1},$$
which produces the discrete equations \eqref{sk-d4}-\eqref{sk-d10}.

As a conclusion of this section, we give the following theorem:
\begin{thm}
The system \eqref{sk-d1}-\eqref{sk-d3} is an integrable
discretization of the extended SK equation \eqref{sk11}. With $v_n= (ln
f_{n})_x$, $ u_n=v_{n,x}$, $ p_n= u_{n,x}$, $ q_n= p_{n,x}$, $ r_n=
q_{n,x}$, $s_n= r_{n,x}$, $\eta_n= s_{n,x}$,
\eqref{sk-d1}-\eqref{sk-d3} can be transformed to
\eqref{sk-d4}-\eqref{sk-d10}, which converge to the extended SK equation \eqref{sk-e1}-\eqref{sk-e3}when
we take a continuum limit.
\end{thm}

\section{Ito Equation}
\setcounter{equation}{0}
In this section, we are going to discretize
the extended Ito equation
\begin{eqnarray}
&& w_{tt}+p_{t}+6uv_{t}=0,\label{ix-e1}\\
&& v_t= w_{xt}, p_t= u_{xt}, u=v_x.\label{ix-e2}
\end{eqnarray}
The dependent variable transformation
\begin{equation}
w=ln f,v=(ln f)_{x},
\end{equation}
gives the bilinear form
\begin{equation}
 D_t(D_t+D_x^3)f \cdot f=0,\label{i11}
\end{equation}
which we also call it Ito equation without confusion. A bilinear
B\"acklund transformation of \eqref{i11} is \cite{ito1}\cite{ito2}
\begin{eqnarray}
&& (D_xD_t-\mu D_x-\gamma D_t+\gamma \mu)f \cdot g=0,\label{i-b1}\\
&& (D_t+3\gamma^2D_x-3\gamma D_x^2+D_x^3-\lambda)f \cdot g=0,\label{i-b2}
\end{eqnarray}
where $\lambda$, $\gamma$ and $\mu$ are arbitrary constants. Setting
$\mu=0$, $\lambda=0$, $\gamma=\frac{2}{h}$, $f\rightarrow f_{n}$ and
$g\rightarrow f_{n-h}$ in \eqref{i11}, \eqref{i-b1} and
\eqref{i-b2}, we get an integrable differential-difference system
\begin{eqnarray}
&& (D_t^2+D_x^3D_t)f_n \cdot f_n=0,\label{ix-d1}\\
&& (D_xD_t-\frac{2}{h} D_t)f_{n} \cdot f_{n-h}=0,\label{ix-d2}\\
&& (hD_t+hD_x^3+\frac{12}{h}D_x-6D_x^2)f_{n} \cdot
f_{n-h}=0.\label{ix-d3}
\end{eqnarray}
Considering $x$ as an auxiliary variable, we next prove that when
$h\rightarrow 0$, the nonlinear form of \eqref{ix-d1}-\eqref{ix-d3}
converges to the extended Ito equation \eqref{ix-e1}-\eqref{ix-e2}. With $ w_n=ln(f_n)$, $v_n=
w_{n,x}$, $ u_n=v_{n,x}$, $ p_n= u_{n,x}$, $q_n=p_{n,x}$, equations
\eqref{ix-d1}-\eqref{ix-d3} can be transformed to
\begin{eqnarray}
&& w_{n,tt}+p_{n,t}+6u_nv_{n,t}=0,\label{ix-d4}\\
&& (v_{n+h,t}+v_{n,t})+( w_{n+h,t}- w_{n,t})
(v_{n+h}-v_{n})=\frac{2}{h}( w_{n+h,t}- w_{n,t}),\label{ix-d5}\\
&& (p_{n+h,t}+p_{n,t})+3( u_{n+h,t}- u_{n,t})
(v_{n+h}-v_{n})+( w_{n+h,t}- w_{n,t})
(p_{n+h}-p_{n})\nonumber\\
&&\quad =\frac{2}{h}( u_{n+h,t}- u_{n,t}),\label{ix-d6}\\
&&6\big(( u_{n+h}+ u_{n})+(v_{n+h}-v_{n})^2\big)=\frac{12}{h}(v_{n+h}-v_{n})
+h(w_{n+h,t}-w_{n,t})\nonumber\\
&&\quad +h\big((p_{n+h}-p_{n})+3(v_{n+h}-v_{n})(u_{n+h}
+u_{n})+(v_{n+h}-v_{n})^3\big).\label{ix-d7}
\end{eqnarray}
Replacing $u_n$ by $u(x,t)$, $u_{n+h}$ by $u(x+h,t)$, and similarly
for $ w$, $v$, $ p$, as $h\rightarrow0$, equations
\eqref{ix-d4}-\eqref{ix-d7} become
\begin{eqnarray}
&& w_{tt}+p_{t}+6uv_{t}=0,\\
&& v_t= w_{xt}, p_t= u_{xt}, u=v_x.
\end{eqnarray}
Thus we have
\begin{thm}
The system \eqref{ix-d1}-\eqref{ix-d3} form an integrable
discretization of the Ito equation \eqref{i11}. With $ w_n=ln(f_n)$,
$v_n= w_{n,x}$, $ u_n=v_{n,x}$, $ p_n= u_{n,x}$, $q_n=p_{n,x}$,
\eqref{ix-d1}-\eqref{ix-d3} can be transformed to
\eqref{ix-d4}-\eqref{ix-d7}, which converge to the extended Ito equation \eqref{ix-e1}-\eqref{ix-e2}
 as $h\rightarrow 0$.
\end{thm}

For the bilinear differential-difference equations
\eqref{ix-d1}-\eqref{ix-d3}, we have the proposition
\begin{prop}
The bilinear equations \eqref{ix-d1}-\eqref{ix-d3} have the
B\"acklund transformation
\begin{eqnarray}
&& (D_x e^{-\frac{h}{2}D_n}+\lambda D_x e^{\frac{h}{2}D_n}-\frac{2}{h}\lambda e^{\frac{h}{2}D_n}+\frac{2}{h} e^{-\frac{h}{2}D_n})f_n \cdot g_n=0,\label{dix-b1}\\
&& (D_t e^{-\frac{h}{2}D_n}-\lambda D_t e^{\frac{h}{2}D_n}-\lambda \omega e^{\frac{h}{2}D_n}+\omega e^{-\frac{h}{2}D_n})f_n \cdot g_n=0,\label{dix-b2}\\
&& (D_t+D_x^3)f_n \cdot g_n=0,\label{dix-b3}\\
&& (D_xD_t-\mu D_x)f_n\cdot g_n=0,\label{dix-b4}
\end{eqnarray}
where $\lambda$, $\omega$, $\mu$ are arbitrary constants.
\end{prop}
{\bf Proof.} Let $f_n$ be a solution of \eqref{ix-d1}-\eqref{ix-d3}
and $g_n$ be given by \eqref{dix-b1}-\eqref{dix-b4}. If we can show
that
\begin{eqnarray*}
&& P_1\equiv (D_t^2+D_x^3D_t)g_n \cdot g_n=0,\\
&& P_2\equiv (D_xD_t-\frac{2}{h} D_t)e^{\frac{h}{2}D_n}g_n \cdot g_{n}=0,\\
&& P_3\equiv (hD_t+hD_x^3+\frac{12}{h}D_x-6D_x^2)e^{\frac{h}{2}D_n}g_n \cdot g_{n}=0,
\end{eqnarray*}
then \eqref{dix-b1}-\eqref{dix-b4} form a B\"acklund transformation
of \eqref{ix-d1}-\eqref{ix-d3}. Noticing that equations
\eqref{i-b1}-\eqref{i-b2} form a B\"acklund transformation of the
Ito equation $(D_t^2+D_x^3D_t)f \cdot f=0$, it is obvious that
$P_1=0$. By using \eqref{dix-b1}-\eqref{dix-b2} and the bilinear
identities \eqref{a1}, \eqref{a8}-\eqref{a10}, we can calculate that
\begin{eqnarray*}
&&\quad -(e^{\frac{h}{2}D_n}f_n\cdot f_n)P_2\\
&&\equiv [(D_xD_t-\frac{2}{h} D_t)e^{\frac{h}{2}D_n}f_n \cdot f_n](e^{\frac{h}{2}D_n}g_n\cdot g_n)\nonumber\\
&&\quad-(e^{\frac{h}{2}D_n}f_n\cdot f_n)[(D_xD_t-\frac{2}{h} D_t)e^{\frac{h}{2}D_n}g_n \cdot g_n]\\
&&=\frac{1}{2}D_x[(D_te^{-\frac{h}{2}D_n}f_n\cdot g_n-\lambda D_te^{\frac{h}{2}D_n}f_n\cdot g_n)\cdot (e^{\frac{h}{2}D_n}f_n\cdot g_n-\lambda^{-1}e^{-\frac{h}{2}D_n}f_n\cdot g_n)]\\
&&=0.
\end{eqnarray*}
By \eqref{dix-b1}, \eqref{dix-b3} and the bilinear identities
\eqref{a1}, \eqref{a2}, \eqref{a6},
\begin{eqnarray*}
&&\quad -(e^{\frac{h}{2}D_n}f_n\cdot f_n)P_3\\
&&\equiv [(D_t+\frac{12}{h^2}D_x-\frac{6}{h} D_x^2+D_x^3)e^{\frac{h}{2}D_n}f_n \cdot f_n)](e^{\frac{h}{2}D_n}g_n\cdot g_n)\\
&&\quad -(e^{\frac{h}{2}D_n}f_n\cdot f_n)[(D_t+\frac{12}{h^2}D_x-\frac{6}{h} D_x^2+D_x^3)e^{\frac{h}{2}D_n}g_n \cdot g_n]\\
&&=-3D_x[(D_x e^{\frac{h}{2}D_n}f_n\cdot g_n-\frac{2}{h}e^{\frac{h}{2}D_n}f_n\cdot g_n)\cdot(D_x e^{-\frac{h}{2}D_n}f_n\cdot g_n+\frac{2}{h}e^{-\frac{h}{2}D_n}f_n\cdot g_n)]\\
&&=0.
\end{eqnarray*}
Thus we have completed the proof.

From the B\"acklund transformation \eqref{dix-b1}-\eqref{dix-b4}, we
can derive a Lax pair for the discrete integrable system
\eqref{ix-d4}-\eqref{ix-d7}. Setting $\mu=-\omega$, $v_n=(ln
g_n)_x$, $ u_n=v_{n,x}$, $p_n=u_{n,x}$, $f_n=\phi_ng_n$,
$\psi_n=\phi_{n,x}$, $\Delta_n=\psi_{n,x}$ and
$\Sigma_n=\Delta_{n,x}$ in \eqref{dix-b1}-\eqref{dix-b4}, we obtain

\begin{displaymath}
L_{1,n}
\left( \begin{array}{c}
\phi_{n+1}\\
\psi_{n+1}\\
\Delta_{n+1}\\
\Sigma_{n+1}
\end{array} \right)=L_{2,n}
\left( \begin{array}{c}
\phi_{n}\\
\psi_{n}\\
\Delta_n\\
\Sigma_n
\end{array} \right)
\end{displaymath}
and
\begin{displaymath}
\left( \begin{array}{c}
\phi_{n,t}\\
\psi_{n,t}\\
\Delta_{n,t}\\
\Sigma_{n,t}
\end{array} \right) =
Q_n
\left( \begin{array}{c}
\phi_{n}\\
\psi_{n}\\
\Delta_n\\
\Sigma_n
\end{array} \right),
\end{displaymath}
where
\begin{displaymath}
L_{1,n}=
\left(
\begin{array}{cccc}
 -\frac{2 \lambda }{h}-\lambda  v_{n}+\lambda  v_{n+1} &  \lambda  &  0 &  0 \\
 -\lambda  u_{n}+\lambda  u_{n+1} &  -\frac{2 \lambda }{h}-\lambda  v_{n}+\lambda  v_{n+1} &  \lambda  &  0 \\
 -\lambda  p_{n}+\lambda  p_{n+1} &  -2 \lambda  u_{n}+2 \lambda  u_{n+1} &  -\frac{2 \lambda }{h}-\lambda  v_{n}+\lambda  v_{n+1} &  \lambda  \\
 -\lambda  \omega +\lambda  w_{n,t}-\lambda  w_{n+1,t} &  6 \lambda  u_{n+1} &  0 &  \lambda  \\
\end{array}
\right),
\end{displaymath}
\begin{displaymath}
L_{2,n}=
\left(
\begin{array}{cccc}
 -\frac{2}{h}-v_{n}+v_{n+1} &  -1 &  0 &  0 \\
 -u_{n}+u_{n+1} &  -\frac{2}{h}-v_{n}+v_{n+1} &  -1 &  0 \\
 -p_{n}+p_{n+1} &  -2 (u_{n}-u_{n+1}) &  -\frac{2}{h}-v_{n}+v_{n+1} &  -1 \\
 -\omega -w_{n,t}+w_{n+1,t} &  6 u_{n} &  0 &  1 \\
\end{array}
\right),
\end{displaymath}
\begin{displaymath}
Q_n =
\left(
\begin{array}{cccc}
 0 & -6 u_{n} & 0 & -1 \\
 -2 v_{n,t} & -\omega  & 0 & 0 \\
 -2 u_{n,t} & -2 v_{n,t} & -\omega  & 0 \\
 -2 p_{n,t} & -4 u_{n,t} & -2 v_{n,t} & -\omega  \\
\end{array}
\right).
\end{displaymath}
The compatibility condition of the two linear problems above is
$$L_{2,n,t}L_{2,n}^{-1}-L_{1,n,t}L_{1,n}^{-1}=L_{1,n}Q_{n+1}L_{1,n}^{-1}-L_{2,n}Q_{n}L_{2,n}^{-1},$$
which gives the discrete equations \eqref{ix-d4}-\eqref{ix-d7}.

\section{Conclusion and Discussion}
In this paper, based on the bilinear method, we have presented a
systematic procedure towards finding integrable discretizations. The
key to this method focuses on deriving compatible equations. Here we
have chosen the well-known B\"acklund transformation. We must remark
that generating difference equations from B\"acklund transformations
has been studied for many years.  Only with the convergence
condition ($D_x=D_n+O(h^k)$, $k\geq1$ for the $x$-direction
discretization) concerned, the difference equations may converge to
the original equation and thus approach the integrable
discretization.

This procedure would seem to merit further attention. Our plans include conducting
numerical studies, and trying to work out a suitable inverse scattering formalism.
Full discretization of soliton equations and the promotion to other
kinds of equations, such as the NLS equation or the Lotka-Voterra
system will also be considered.

\section{Acknowledgements}
Y N Zhang is grateful to Decio Levi for helpful discussions. This work of H W Tam is  supported by the Hong Kong Baptist University Faculty Research
Grant FRG2/11-12/065 and the Hong Kong Research Grant Council under grant GRF HKBU
202512. Y N Zhang, X K Chang and X B Hu are supported by the National Natural Science Foundation of China (grant no.11331008). J Hu is supported by the National Natural Science Foundation of China (grant no.11201425).

\appendix
  \renewcommand{\thesection}{Appendix \Alph{section}}
\setcounter{equation}{0}
\renewcommand{\theequation}{A.\arabic{equation}}
\section{Bilinear Operator Identities}

\begin{eqnarray}
&&\quad (D_xe^{\frac{h}{2}D_n}f_n \cdot f_n)(e^{\frac{h}{2}D_n}g_n\cdot g_n)-(e^{\frac{h}{2}D_n}f_n\cdot f_n)(D_xe^{\frac{h}{2}D_n}g_n \cdot g_n)\nonumber\\
&& =2\sinh(\frac{h}{2}D_n)(D_xf_n\cdot g_n)\cdot(f_ng_n)\nonumber\\
&& =D_x(e^{\frac{h}{2}D_n}f_n\cdot g_n)\cdot(e^{-\frac{h}{2}D_n}f_n\cdot g_n).\label{a1}
\end{eqnarray}
\begin{eqnarray}
&&\quad (D_x^2e^{\frac{h}{2}D_n}f_n\cdot f_n)(e^{\frac{h}{2}D_n}g_n\cdot g_n)-(e^{\frac{h}{2}D_n}f_n\cdot f_n)(D_x^2e^{\frac{h}{2}D_n}g_n \cdot g_n)\nonumber\\
&&=D_x[(D_xe^{\frac{h}{2}D_n}f_n\cdot g_n)\cdot(e^{-\frac{h}{2}D_n}f_n\cdot g_n)\nonumber\\
&&\quad-(e^{\frac{h}{2}D_n}f_n\cdot g_n)\cdot(D_x^{-\frac{h}{2}D_n}f_n\cdot g_n)].\label{a2}
\end{eqnarray}
\begin{eqnarray}
&&\quad D_x[(D_xe^{\frac{h}{2}D_n}f_n\cdot g_n)\cdot(e^{-\frac{h}{2}D_n}f_n\cdot g_n)+(e^{\frac{h}{2}D_n}f_n\cdot g_n)\cdot(D_x^{-\frac{h}{2}D_n}f_n\cdot g_n)]\nonumber\\
&& =2\sinh(\frac{h}{2}D_n)(D_x^2f_n\cdot g_n)\cdot(f_ng_n).\label{a3}
\end{eqnarray}
\begin{eqnarray}
&&\quad (D_xD_ye^{\frac{h}{2}D_n}f_n \cdot f_n)(e^{\frac{h}{2}D_n}g_n\cdot g_n)-(e^{\frac{h}{2}D_n}f_n\cdot f_n)(D_xD_ye^{\frac{h}{2}D_n}g_n \cdot g_n)\nonumber \\
&&=D_y[(D_xe^{\frac{h}{2}D_n}f_n\cdot g_n)\cdot (e^{-\frac{h}{2}D_n}f_n\cdot g_n)-
(e^{\frac{h}{2}D_n}f_n\cdot g_n)\cdot (D_xe^{-\frac{h}{2}D_n}f_n\cdot g_n)]\nonumber\\
&&\quad +(D_xe^{\frac{h}{2}D_n}f_n\cdot f_n)\cdot (D_ye^{\frac{h}{2}D_n}g_n\cdot g_n)\nonumber\\
&&\quad-(D_ye^{\frac{h}{2}D_n}f_n\cdot f_n)\cdot (D_xe^{\frac{h}{2}D_n}g_n\cdot g_n).\label{a4}
\end{eqnarray}
\begin{eqnarray}
&&\quad D_y[(D_xe^{\frac{h}{2}D_n}f_n\cdot g_n)\cdot(e^{-\frac{h}{2}D_n}f_n\cdot g_n)+(e^{\frac{h}{2}D_n}f_n\cdot g_n)\cdot(D_x^{-\frac{h}{2}D_n}f_n\cdot g_n)]\nonumber\\
&&=2\sinh(\frac{h}{2}D_n)[(D_xD_yf_n\cdot g_n)\cdot(f_ng_n)+(D_yf_n\cdot g_n)\cdot(D_xf_n\cdot g_n)].\label{a5}
\end{eqnarray}
\begin{eqnarray}
&&\quad (D_x^3e^{\frac{h}{2}D_n}f_n \cdot f_n)(e^{\frac{h}{2}D_n}g_n\cdot g_n)-(e^{\frac{h}{2}D_n}f_n\cdot f_n)(D_x^3e^{\frac{h}{2}D_n}g_n \cdot g_n)\nonumber\\
&&=2\sinh(\frac{h}{2}D_n)(D_x^3f_n\cdot g_n)\cdot(f_ng_n)\nonumber\\
&&\quad-3D_x(D_xe^{\frac{h}{2}D_n}f_n\cdot g_n)\cdot(D_xe^{-\frac{h}{2}D_n}f_n\cdot g_n).\label{a6}
\end{eqnarray}
\begin{eqnarray}
&&\quad (D_xe^{\frac{h}{2}D_n}f_n \cdot f_n)(D_x^2e^{\frac{h}{2}D_n}g_n\cdot g_n)-(D_x^2e^{\frac{h}{2}D_n}f_n\cdot f_n)(D_xe^{\frac{h}{2}D_n}g_n \cdot g_n)\nonumber\\
&&=2\sinh(\frac{h}{2}D_n)(D_xf_n\cdot g_n)\cdot(D_x^2f_n\cdot g_n)\nonumber\\
&&\quad+D_x(D_xe^{\frac{h}{2}D_n}f_n\cdot g_n)\cdot(D_xe^{-\frac{h}{2}D_n}f_n\cdot g_n)\nonumber\\
&&=-D_x[(D_x^2e^{\frac{h}{2}D_n}f_n\cdot g_n)\cdot(e^{-\frac{h}{2}D_n}f_n\cdot g_n)\nonumber\\
&&\quad +(e^{\frac{h}{2}D_n}f_n\cdot g_n)\cdot(D_x^2e^{-\frac{h}{2}D_n}f_n\cdot g_n)\nonumber\\
&&\quad +(D_xe^{\frac{h}{2}D_n}f_n\cdot g_n)\cdot(D_xe^{-\frac{h}{2}D_n}f_n\cdot g_n)].\label{a7}
\end{eqnarray}
\begin{eqnarray}
&&\quad (D_zD_te^{\frac{h}{2}D_n}f_n \cdot f_n)(e^{\frac{h}{2}D_n}g_n\cdot g_n)-(e^{\frac{h}{2}D_n}f_n\cdot f_n)(D_zD_te^{\frac{h}{2}D_n}g_n \cdot g_n)\nonumber \\
&&=\frac{1}{2}D_z[(D_te^{\frac{h}{2}D_n}f_n\cdot g_n)\cdot (e^{-\frac{h}{2}D_n}f_n\cdot g_n)\nonumber\\
&&\quad-(e^{\frac{h}{2}D_n}f_n\cdot g_n)\cdot (D_te^{-\frac{h}{2}D_n}f_n\cdot g_n)]\nonumber\\
&&\quad +\frac{1}{2}D_t[(D_ze^{\frac{h}{2}D_n}f_n\cdot g_n)\cdot (e^{-\frac{h}{2}D_n}f_n\cdot g_n)\nonumber\\
&&\quad-(e^{\frac{h}{2}D_n}f_n\cdot g_n)\cdot (D_ze^{-\frac{h}{2}D_n}f_n\cdot g_n)].\label{a8}
\end{eqnarray}
\begin{eqnarray}
&&\quad D_t[(D_ze^{-\frac{h}{2}D_n}f_n\cdot g_n)\cdot(e^{-\frac{h}{2}D_n}f_n\cdot g_n)]\nonumber\\
&&=D_z[(D_te^{-\frac{h}{2}D_n}f_n\cdot g_n)\cdot(e^{-\frac{h}{2}D_n}f_n\cdot g_n)].\label{a9}
\end{eqnarray}
\begin{eqnarray}
&&\quad D_t[(e^{\frac{h}{2}D_n}f_n\cdot g_n)\cdot(D_ze^{\frac{h}{2}D_n}f_n\cdot g_n)]\nonumber\\
&&=D_z[(e^{\frac{h}{2}D_n}f_n\cdot g_n)\cdot(D_te^{\frac{h}{2}D_n}f_n\cdot g_n)].\label{a10}
\end{eqnarray}
\begin{eqnarray}
&&\quad (D_x^5e^{\frac{h}{2}D_n}f_n \cdot f_n)(e^{\frac{h}{2}D_n}g_n\cdot g_n)-(e^{\frac{h}{2}D_n}f_n\cdot f_n)(D_x^5e^{\frac{h}{2}D_n}g_n \cdot g_n)\nonumber\\
&&=2\sinh(\frac{h}{2}D_n)[(D_x^5f_n\cdot g_n)\cdot(f_ng_n)+5(D_x^3f_n\cdot g_n)\cdot(D_x^2f_n\cdot g_n)]\nonumber\\
&&\quad -5D_x[(D_x^3e^{\frac{h}{2}D_n}f_n\cdot g_n)\cdot(D_xe^{-\frac{h}{2}D_n}f_n\cdot g_n)\nonumber\\
&&\quad +(D_xe^{\frac{h}{2}D_n}f_n\cdot g_n)\cdot(D_x^3e^{-\frac{h}{2}D_n}f_n\cdot g_n)]\nonumber\\
&&\quad -5[(D_x^3e^{\frac{h}{2}D_n}f_n \cdot f_n)(D_x^2e^{\frac{h}{2}D_n}g_n\cdot g_n)\nonumber\\
&&\quad -(D_x^2e^{\frac{h}{2}D_n}f_n\cdot f_n)(D_x^3e^{\frac{h}{2}D_n}g_n \cdot g_n)].\label{a11}
\end{eqnarray}
\begin{eqnarray}
&&\quad (D_x^4e^{\frac{h}{2}D_n}f_n \cdot f_n)(e^{\frac{h}{2}D_n}g_n\cdot g_n)-(e^{\frac{h}{2}D_n}f_n\cdot f_n)(D_x^4e^{\frac{h}{2}D_n}g_n \cdot g_n)\nonumber\\
&&=D_x[(D_x^3e^{\frac{h}{2}D_n}f_n\cdot g_n)\cdot(e^{-\frac{h}{2}D_n}f_n\cdot g_n)-(e^{\frac{h}{2}D_n}f_n\cdot g_n)\cdot(D_x^3e^{-\frac{h}{2}D_n}f_n\cdot g_n)\nonumber\\
&&\quad -3(D_x^2e^{\frac{h}{2}D_n}f_n\cdot g_n)\cdot(D_xe^{-\frac{h}{2}D_n}f_n\cdot g_n)\nonumber\\
&&\quad +3(D_xe^{\frac{h}{2}D_n}f_n\cdot g_n)\cdot(D_x^2e^{-\frac{h}{2}D_n}f_n\cdot g_n)]\nonumber\\
&&\quad -2[(D_x^3e^{\frac{h}{2}D_n}f_n \cdot f_n)(D_xe^{\frac{h}{2}D_n}g_n\cdot g_n)\nonumber\\
&&\quad -(D_xe^{\frac{h}{2}D_n}f_n\cdot f_n)(D_x^3e^{\frac{h}{2}D_n}g_n \cdot g_n)].\label{a12}
\end{eqnarray}

%

\end{document}